\documentclass[12pt]{article}  %envcountsame
\usepackage{amsmath}
\usepackage{amsfonts,amssymb}
\newcommand{\cH}{{\mathcal H}}
\newcommand{\cF}{{\mathcal F}}
\newcommand{\cM}{{\mathcal M}}

\newcommand{\cP}{{\mathcal P}}
\newcommand{\cE}{{\mathcal E}}
\newcommand{\cA}{{\mathcal A}}

\newcommand{\cN}{{\mathcal N}}

\def\R{{\mathbb R}}

\def\Z{{\mathbb Z}}
\def\C{{\mathbb C}}

\newcommand{\1}{{\bf 1}}

\newtheorem{theorem}{Theorem}[section]

\newtheorem{proposition}{Proposition}[section]
\newtheorem{lemma}{Lemma}[section]
\newtheorem{corollary}{Corollary}[section]

\newtheorem{remark}{Remark}[section]

\newenvironment{proof}{\bigskip\par\noindent{\it Proof:}}{$\square$\newline\vspace*{0.2cm}}

\makeatletter
    
    \@addtoreset{equation}{section}
\makeatother

\makeatother 
\begin{document}

\title{{Ergodic Property of Markovian Semigroups on Standard Forms
 of von Neumann Algebras}}
%\titlerunning{Structure of Dirichlet forms}

\author{Yong Moon Park}
\date{
  { \small Department of Mathematics, Yonsei University\\
   Seoul 120-749, Korea \\
 E-mail : ympark@yonsei.ac.kr }}
%\institute{Department of Mathematics and Institute for Mathematical Sciences,\\
 % Yonsei University, Seoul 120-749, Korea \\
 %\email{ympark@yonsei.ac.kr} }
%\authorrunning{Y. M. Park}

%\date{Received:  / Accepted: }

%\communicated{}

\maketitle
\begin{abstract}
We give sufficient conditions for ergodicity of the Markovian
semigroups associated to Dirichlet forms on standard forms of von
Neumann algebras constructed by the method proposed in
Refs.\cite{Par1,Par2}. We apply our result to show that the
diffusion type Markovian semigroups for quantum spin systems are
ergodic in the region of high temperatures where the uniqueness of
the KMS-state holds.

\vspace*{0.3cm} \noindent {\it Keywords} : Standard forms of von
Neumann Algebras; Dirichlet forms; Markovian semigroups;
ergodicity; quantum spin systems; KMS-states.
\end{abstract}
\baselineskip=24pt

\section{Introduction}
The purpose of this work is to investigate ergodic property of the
Markovian semigroups associated to Dirichlet forms on the standard
form of a von Neumann algebra $\cM$ acting on a Hilbert space
$\cH$ with a cyclic and separating vector $\xi_0$. Denote by
$\sigma_t,\,t\in\R$, the modular automorphism on $\cM$ associated
with the pair $(\cM,\xi_0)$ and $\cM_{1/2}$ the dense subset of
$\cM$ consisting of every $\sigma_t$-analytic element on a domain
containing the strip $\{ z\in\mathbb{C} : |\text{Im} z|\le
1/2\}$\cite{BR}. Let $\{x_k: k\in I\}$ be a (finite or countable)
family of elements in $\cM_{1/2}$ which generates $\cM$. Let
$(\cE,D(\cE))$ be the Dirichlet form constructed with $\{x_k: k\in
I\}$ and an admissible function by means of Refs.
\cite{Par1,Par2}. For the details, see Section 2. Denote by
$T_t,\,t\ge0$, the Markovian semigroup associated to
$(\cE,D(\cE))$. Let $\mathcal{N}$ be the fixed point space of
$T_t$;
$$\mathcal{N}=\{\eta\in\cH:T_t\eta=\eta,\,\forall t\ge0\}.$$
We show that $\mathcal{N}=[\mathcal{Z}(\cM)\xi_0]$, where
$\mathcal{Z}(\cM)$ is the center of $\cM$;
$\mathcal{Z}(\cM)=\cM\cap\cM'$, and $[\mathcal{Z}(\cM)\xi_0]$ is
the closure of $\mathcal{Z}(\cM)\xi_0$. As a consequence, $T_t$ is
ergodic if and only if $\cM$ is a factor. We apply our result to
the translation invariant Markovian semigroups for quantum spin
systems \cite{Par1}, and show that the semigroups are ergodic in
region of high temperatures where the uniqueness of the KMS-state
holds.

Let us describe the background of this study briefly. The need to
construct Markovian semigroups on von Neumann algebras, which are
(KMS) symmetric with respect to non-tracial states, is clear for
various applications to open systems\cite{Dav}, quantum
statistical mechanics\cite{BR} and quantum
probability\cite{Acc,Part}. Although on the abstract level we have
quite well-developed theory\cite{Cip1,GL1,GL2}, the progress in
concrete applications is relatively slow. For construction of
Dirichlet forms and associated Markovian semigroups, we refer to
\cite{BKP1,BKP2,MZ1,MZ2,Par1,Zeg} and the references there in.

During the last ten years, systematic methods to construct
Dirichlet forms and associated Markovian semigroups of jump and
diffusion types have been developed. Nontrivial translation
invariant symmetric semigroups of jump type for quantum spin
systems have been constructed and the strong ergodicity of the
semigroups has been established in Refs. \cite{MZ1,MZ2}. See also
\cite{Zeg} and the references there in. In \cite{Par1}, we gave a
general construction method of Dirichlet forms of diffusion type
in the framework of the general theory of Dirichlet forms and
Markovian semigroups on standard forms of von Neumann algebras
developed by Cipriani\cite{Cip1}. The method has been used
successfully to construct Dirichlet forms and associated Markovian
semigroups for quantum spin systems \cite{Par1}, CCR and CAR
algebras with respect to quasi-free states\cite{BKP1,BKP2,KP}, and
quantum mechanical systems\cite{BK}. Recently, in \cite{Par2} we
have shown that the symmetric embedding of a general Lindblad type
(bounded) generator of a quantum dynamical semigroup satisfying
KMS-symmetry can be written in the form of a Dirichlet operator
associated to a Dirichlet form given in \cite{Par1}.

The next step in this research area would be the investigation of
detailed properties of Markovian semigroups, such as ergodicity,
mixing property and convergence to the equilibrium, etc. In the
case of CCR and CAR algebras with respect to quasi-free states,
the spectrum of the generators of the Markovian semigroups
constructed in \cite{CFL,BKP1,BKP2,KP} has been analyzed. However,
in general the detail properties of the Markovian semigroups
associated to Dirichlet forms in \cite{Par1,Par2} are hard to be
established. Thus it would be nice to have a simple criteria
whether the Markovian semigroup under study is ergodic or not in
the sense of Cipriani\cite{Cip2}.

We organize the paper as follows: In Section 2, we introduce
notations and terminologies, and then list our results(Theorem
\ref{thm2.1} and Corollary \ref{cor2.1} ). We also give comments
(Remark \ref{rem2.1}) on possible applications of our results.
Section 3 is devoted to the proof of Theorem \ref{thm2.1}. We
first describe the basic ideas of the proof and then establish
some technical lemmas(Lemma \ref{lemm3.1} - Lemma \ref{lemm3.5})
which are need in the sequel. Using the lemmas we complete the
proof of Theorem \ref{thm2.1}. In Section 4, we apply our main
results to prove that the diffusion type translation invariant
Markovian semigroups for quantum spin systems constructed in
\cite{Par1} are ergodic in the region of high temperature where
the uniqueness of the KMS state holds.

\section{ Notation, Terminologies and Main Results}

We first  introduce necessary terminologies in the theory of
Dirichlet forms and Markovian semigroups on standard form of von
Neumann algebras\cite{Cip1}. Next we give a brief review on the
construction of Dirichlet forms developed in \cite{Par1} and then
list our main results.

Let $\cM$ be a $\sigma$-finite von Neumann algebra
 acting on a complex Hilbert space $\cH$ with an inner product $\langle \cdot , \cdot \rangle$
 which is conjugate linear in the first and linear in the second variable.
 Let $\xi_0$ be a cyclic and separating vector
 for $\cM$. We use $\Delta$ and  $J$ to denote respectively, the modular operator and
 the modular conjugation associated with the pair $(\cM,\, \xi_0 )$\cite{BR}. The associated
 modular automorphism is denoted by $\sigma_t$: $\sigma_t (A) = \Delta^{it} A \Delta^{-it},
  A\in \mathcal{M}$. Finally, $j:\mathcal{M}\rightarrow\mathcal{M}'$ is the antilinear
 $*$-isomorphism defined by $j(A)=JAJ,\,A\in\cM$, where $\cM'$ is
 the commutant of $\cM$. By the Tomita-Takesaki theorem(Theorem
 2.5.14 of \cite{BR}), it follows that $\sigma_t(\cM)=\cM$ and $j(\cM)=\cM'.$

 The natural positive cone $\mathcal{P}$ associated with the pair
  $(\mathcal{M}$, $\xi_0 )$ is the closure of the set
 $$
 \{ Aj(A)\xi_0 : A\in \mathcal{M}\}.
 $$
By a general result, the closed convex cone $\cP$ can be obtained
by the closure of the set
 $$
 \{\Delta^{1/4} AA^* \xi_0 : A\in \mathcal{M}\}
 $$
and this cone $\cP$ is
  self-dual  in the sense  that
$$\Big\{\xi \in \cH : \langle\xi, \eta\rangle \ge 0,\,\,\forall \eta \in \cP\Big\}=\cP.$$
For the details we refer to \cite{Ara} and Section 2.5 of
\cite{BR}.

   The form
$(\mathcal{M},\mathcal{H},\mathcal{P},J)$ is the standard form
associated with the pair $(\mathcal{M},\xi_0 )$.  We shall use the
fact that $\cH$ is  the complexification
 of the real  subspace
 $ \cH^J =\Big\{\xi\in\cH : \langle\xi, \eta\rangle\in\mathbb{R},\,\,\forall\eta\in\cP\Big\},$
 whose elements are called {\it $J$-real}: $\cH =\cH^J \bigoplus i \cH^J .$
  The cone $\cP$ gives rise to a structure of ordered Hilbert space on $\cH^J$
 (denoted by $\leq$) and to an anti-linear involution $J$ on $\cH$,
 which preserves $\cP$ and $\cH^J$: $J(\xi +i\eta) = \xi -i\eta,\,
 \forall \xi, \eta \in \cH^J$.  Also note that any $J$-real element $\xi \in \cH^J$ can be decomposed uniquely
 as a difference  of two mutually orthogonal, positive and  negative part of $\xi$, respectively :
  $\xi =\xi_+ - \xi_-,\quad \xi_+,\,\xi_- \in \cP$
 and $\langle\xi_+, \xi_- \rangle =0$.

A bounded operator $A$ on $\cH$ is called {\it $J$-real} if
$AJ=JA$ and {\it positive preserving} if $A\cP \subset \cP$. A
semigroup $\{T_t\}_{t\ge 0}$ is said to be {\it $J$-real} if $T_t$
is $J$-real for any $t\ge0$ and it is called {\it positive
preserving} if $T_t$ is  positive preserving for any $t\ge 0$.
 A bounded operator $A:\cH \to \cH$ is called {\it sub-Markovian}
(with respect to $\xi_0$) if $0\le \xi\le \xi_0$ implies $0\le
A\xi \le \xi_0$.  $A$
 is called {\it Markovian} if it is sub-Markovian and also
 $A\xi_0=\xi_0$. A semigroup $\{T_t\}_{t\ge 0}$ is said to be {\it
 sub-Markovian} (with respect to $\xi_0$) if $T_t$ is sub-Markovian for
 every $t\ge 0$. A semigroup $\{ T_t\}_{t\ge 0}$ is called {\it
 Markovian} if $T_t$ is Markovian for every $t\ge 0$.

Next, we consider a sesquilinear form on some linear manifold of
$\cH$ : $\cE(\cdot,\cdot): D(\cE)\times D(\cE) \to \C$. We also
consider the associated quadratic form: $\cE[\cdot]: D(\cE)\to
\C$,  $\cE[\xi]:= \cE(\xi,\xi)$.  A real valued quadratic form
$\cE[\cdot]$ is said to be {\it semi-bounded}(or {\it  bounded
below}) if $\inf \{ \cE[\xi] : \xi \in D(\cE), \,\, ||\xi|| =1\}
=-b > -\infty$. A quadratic form $(\cE, D(\cE))$ is said to be
{\it $J$-real} if $J D(\cE) \subset D(\cE)$ and $\cE[J\xi] =
\overline{\cE[\xi]}$ for any $\xi \in D(\cE)$. For a semi-bounded
quadratic form $\cE$, one considers the inner product given by
$\langle\xi,\eta\rangle_\lambda:= \cE(\xi,\eta) + \lambda
\langle\xi,\eta\rangle$, for $\lambda>b$. The form $(\cE,D(\cE))$
is {\it closed} if $D(\cE)$ is a Hilbert space for some of the
above inner products. The form $(\cE,D(\cE))$ is called {\it
closable} if it admits a closed extension.

Associated to a semi-bounded closed form $(\cE,D(\cE))$, there are
a self-adjoint operator $(H, D(H))$ and a strongly continuous,
symmetric semigroup $\{T_t\}_{t\ge0}$. Each of the above objects
determines uniquely the others according to well known relations
(see \cite{RS} and Section 3.1 of \cite{BR}).

Let us denoted by $Proj(\xi,Q)$ the projection of the vector
$\xi\in\cH^J$ onto the closed, convex cone $Q\subset\cH^J$. For
$\xi,\eta\in\cH^J$, define
\begin{eqnarray*}
&&\xi\vee\eta:=Proj(\xi,\eta+\cP),\\
&&\xi\wedge\eta:=Proj(\xi,\eta-\cP).
\end{eqnarray*}
A $J$-real, real-valued, densely defined quadratic form $(\cE,
D(\cE))$ is called {\it Markovian} with respect to $\xi_0\in \cP$
if
\begin{equation*}
\xi \in D(\cE)^J \text{  implies  } \xi\wedge\xi_0 \in D(\cE)
\text{ and } \cE[\xi\wedge\xi_0] \le \cE[\xi],
\end{equation*}
where $D(\cE) ^J := D(\cE) \cap \cH^J$. A closed Markovian form is
called a {\it Dirichlet form}.

Next, we collect main results of \cite{Cip1}. Let $(\cE, D(\cE))$
be a $J$-real, real valued, densely defined closed form. Assume
that the following properties hold:
\begin{eqnarray}\label{2.1}
             &(a)& \xi_0 \in D(\cE), \\
             &(b)&  \cE(\xi,\xi)\ge0 \,\,\text{for} \,\,\xi\in D(\cE), \nonumber \\
             &(c)& \xi\in D(\cE)^J \,\text{ implies } \, \xi_{\pm} \in D(\cE) \,\,\text{
             and}\,\,
 \cE(\xi_+ , \xi_{-} )\leq 0.\nonumber
    \end{eqnarray}
    Then $(\cE, D(\cE))$ is a Dirichlet form if and only if
    $\cE(\xi,\xi_0)\ge0$ for all $\xi\in D(\cE)\cap\cP$.
 The above result follows from Proposition 4.5 (b) and Proposition
4.10 (ii) of \cite{Cip1}. The following is one of main results
(Theorem 4.11) of \cite{Cip1} :
 Let $\{T_t \}_{t\ge 0}$ be a $J$-real, strongly continuous, symmetric semigroup on $\cH$
 and let $(\cE, D(\cE))$ be the  associated densely defined $J$-real, real
 valued quadratic form. Then the followings are equivalent:
 \begin{eqnarray*}
             &(a)& \{ T_t \}_{t\ge 0} \,\, \text{ is
             sub-Markovian}.\\ \nonumber
             &(b)& (\cE, D(\cE))\,\,\text{ is a Dirichlet form .}
    \end{eqnarray*}
We refer the reader to \cite{Cip1} for the details.

Next, we describe the construction of Dirichlet forms developed in
\cite{Par1}. See also \cite{Par2}. For any $\lambda >0$, denote by
$I_\lambda$ the closed strip given by
\begin{equation} \label{2.2}
I_\lambda = \{ z : z \in \C , | Im\, z | \le \lambda \}.
\end{equation}
Recall that an analytic function $f: D \to \mathbb{C}$ on a domain
$D$ containing the strip $I_{1/4}$ is called {\it admissible} if
the following properties hold:
\begin{eqnarray}\label{2.3}
 &(a)& f(t) \ge 0 \quad \text{for}\quad  \forall t\in
 \mathbb{R},\\
&(b)& f(t+i/4) + f(t-i/4) \ge 0 \quad \text{for}\quad \forall t\in
\mathbb{R},\nonumber\\
 &(c)& \text{there exist}\,\, M>0 \,\,\text{and}\,\, p>1\,\,\text{such that the
 bound}\nonumber\\
  && \quad\quad \quad |f(t+is)| \le M(1+|t|)^{-p}\nonumber\\
  &&\text{ holds uniformly in}\,\,s\in [-1/4,1/4].\nonumber
 \end{eqnarray}
We also consider the function $f_0 : \R \to \R$ given by
\begin{equation}\label{2.4}
f_0 (t) = 2 ( e^{2\pi t } + e^{-2 \pi t } ) ^{-1} .
\end{equation}
 One can see that $f_0$ has an analytic extension, denoted by $f_0$ again, to the
interior of $I_{1/4}$.

For any $\lambda >0$, denote by $\cM_\lambda$ the dense subset of
$\cM$ consisting of every $\sigma_t$-analytic element with a
domain containing $I_\lambda$. By Proposition 2.5.21 of \cite{BR},
any $A \in \cM_\lambda$ is strongly analytic on $I_\lambda$. We
denote by $\cM_0$ the dense subset of $\cM$ consisting of every
$\sigma_t$-entire analytic element, i.e.,
$\cM=\bigcap_\lambda\cM_{\lambda}.$

 Let $I$ be a finite or countable (index) set. For
given family $\{x_k: k\in I\}\subset \cM_{1/2}$ of self-adjoint
elements in $\cM_{1/2}$ and an admissible function $f$ or $f=f_0$,
define a sesquilinear form  by
\begin{eqnarray}
     &&D(\cE)=\{\xi\in\cH : \sum_{k\in I}\cE_k(\xi,\xi)<\infty\}, \label{2.5}\\
     &&\cE(\eta,\xi)=\sum_{k\in I}\cE_k(\eta,\xi),\label{2.6}
\end{eqnarray}
where for each $k\in I$
\begin{eqnarray}
     &&\cE_k(\eta,\xi)\label{2.7}\\
     &&\quad=\int \Big\langle\big(\sigma_{t-i/4}(x_k)-j(\sigma_{t-i/4}(x_k))\big)\eta,
          \big(\sigma_{t-i/4}(x_k)-j(\sigma_{t-i/4}(x_k))\big)\xi\Big\rangle f(t)dt\nonumber.
     \end{eqnarray}
For each $k\in I$, the above form is positive and bounded. In
fact, the form $(\cE_k,\cH)$ is a Dirichlet form for each $k\in I$
by Theorem 3.1 of \cite{Par1}. See also Theorem 2.1 of \cite{Par2}
for $f=f_0$. Moreover, if $D(\cE)$ is dense in $\cH$, then the
form $(\cE,D(\cE)$ given in (\ref{2.6}) is a Dirichlet form by
Theorem 5.2 of \cite{Cip1}.

 Before proceeding further, we would like to make a few remarks.
 The function $f_0$ given in (\ref{2.5}) played a special role in \cite{Par2}.
 The symmetric embedding of a general Lindblad type (bounded) generator of a quantum dynamical semigroup
 (satisfying KMS-symmetry) on $\cM$
  can be written as the Dirichlet operator associated to a
  Dirichlet form in (\ref{2.6}) with $f=f_0$. Next, we would like
  to mention that it is not necessary to assume that each $x_k$ in
  $\{x_k :k\in I\}$ is self-adjoint if one defines the Dirichlet
  form $(\cE_k,\cH)$ in (\ref{2.7}) appropriately as in
  (2.6) in \cite{Par2}. Note that, by a simple
  transformation, one can write $\cE_k(\eta,\xi)$ as a sum of two
  Dirichlet forms corresponding two self-adjoint elements. See
  Remark 2.1 (a) in \cite{Par2}. Thus without loss of the
  generality, we assume that each $x_k$ is self-adjoint.

A family $\{x_k :k\in I\}$ is said to {\it generate} $\cM$ if the
$*$-algebra generated by $\{x_k :k\in I\}$ is dense in $\cM$. For
given $\{x_k :k\in I\}\subset\cM_{1/2}$ of self-adjoint elements
and either an admissible function $f$ or else $f=f_0$, let
$(\cE,D(\cE))$ be the Dirichlet form defined as in (\ref{2.5}) -
(\ref{2.7}). Denote by $(H,D(H))$ and $\{T_t\}_{t\ge0}$ the
Dirichlet operator and Markovian semigroup associated to
$(\cE,D(\cE))$, i.e., $T_t=e^{-tH}$. We denote by $\mathcal{N}$
the fixed point space of $T_t$:
\begin{eqnarray}\label{2.8}
\mathcal{N}&:=&\{\eta\in\cH : T_t\eta=\eta\}\\
            &=&\{\eta\in\cH : H\eta=0\}.\nonumber
 \end{eqnarray}
The following is the main result in this paper:
\begin{theorem}\label{thm2.1}
For a family  $\{x_k :k\in I\}\subset\cM_{1/2}$ of self-adjoint
elements and an admissible function $f$ or else $f=f_0$, let
$(\cE,D(\cE))$ be the densely defined Dirichlet form given as in
(\ref{2.5}) -
 (\ref{2.7}). Assume that $\{x_k : k\in I\}$ generates $\cM$. Then
the equality
 $$\mathcal{N}=[\mathcal{Z}(\cM)\xi_0]$$
 holds, where $\mathcal{Z}(\cM)$ is the center of $\cM$, i.e.,
 $\mathcal{Z}(\cM)=\cM\cap\cM'$, and $[\mathcal{Z}(\cM)\xi_0]$ is
 the closure of $\mathcal{Z}(\cM)\xi_0$.
\end{theorem}
Recall that a symmetric, strongly continuous, positive preserving
semigroup $\{T_t\}_{t\ge0}$ on $\cH$ is called {\it ergodic} if
for each $\xi,\eta\in\cP,\,\xi,\eta\neq0$, there exists $t>0$ such
that $\langle\xi,T_t\eta\rangle>0$ \cite{Cip2}. Assume that $\inf
\sigma(H)$ is an eigenvalue  of the generator $H$ of
$\{T_t\}_{t\ge0}$. Then the ergodicity of $\{T_t\}_{t\ge0}$ is
equivalent to that $\inf \sigma(H)$ is a simple eigenvalue of $H$
with a strictly positive (cyclic and separating) eigenvector
(Theorem 4.3 of \cite{Cip2}). As a consequence of Theorem
\ref{thm2.1}, we have the following:
\begin{corollary}\label{cor2.1}
Let $\cM$ be a factor, i.e., $\mathcal{Z}(\cM)=\C\1$. Let
$(\cE,D(\cE))$ be the densely defined Dirichlet form as in Theorem
\ref{thm2.1} and $T_t$ the associated Markovian semigroup. Under
the assumptions as in Theorem \ref{thm2.1}, $\{T_t\}_{t\ge0}$ is
ergodic in the sense that zero is a simple eigenvalue of the
generator $H$ of $T_t$ with eigenvector $\xi_0$.
\end{corollary}
\begin{proof}
Under the assumptions, $\mathcal{N}=\C\xi_0$ by Theorem
\ref{thm2.1}. Since
$j(\sigma_{t-i/4}(x))\xi_0=\sigma_{t-i/4}(x)\xi_0$, it follows
from (\ref{2.7}) that $\cE_k(\xi_0,\xi_0)=0$ for each $k\in I$ and
so $\cE(\xi_0,\xi_0)=0$, which implies that $H\xi_0=0$. See also
Theorem 3.1 (a) of \cite{Par1}. Hence zero is a simple eigenvalue
of $H$ with eigenvector $\xi_0$.
\end{proof}

We will produce the proof of Theorem \ref{thm2.1} in the next
section. Before closing this section, it may be worth to give
comments on possible applications of Theorem \ref{thm2.1}.
\begin{remark}\label{rem2.1}
 (a) In order to apply Theorem \ref{thm2.1} (and Corallary \ref{cor2.1}) to concrete models, one has to choose
a family $\{x_k :k\in I\}\subset\cM_{1/2}$ which generates $\cM$.
Recall that the condition $x_k\in\cM_{1/2}\subset\cM_{1/4}$ for
each $x_k$ is needed for $(\cE_k,\cH)$ to be well defined. If
$\cH$ is a finite dimensional Hilbert space, then the modular
operator $\Delta$ is bounded and so every element $x$ of $\cM$ is
$\sigma_t$-entire analytic. In general, it would be not easy to
choose a generating family $\{x_k : k\subset I\}$ from $\cM_{1/2}$
directly.

(b) For quantum spin systems in the region of high temperatures,
every local observable belongs to $\cM_{1/2}$. In this case, the
choice of $\{x_k: k\in I\}$ is easy. See Section 4 for the
details.

(c) Let $\{f_n : n\in \mathbb{N}\}$ be an orthonormal basis for
$L^2(\R^d)$ and let $a^*(f_n)$ and $a(f_n), \,n\in\mathbb{N},$ be
the creation and annihilation operators which generate a CAR
algebra $\cA$. Let $\omega$ be a quasi-free state on $\cA$ and
$(\cH_{\omega},\pi_{\omega}(\cA),\Omega_{\omega})$ be the cyclic
representation associated to $(\cA,\omega)$. Let
$\cM=\pi_{\omega}(\cA)''$ and $\xi_0=\Omega_{\omega}$. Then for
each $n\in\mathbb{N},$ $\pi_{\omega}(a(f_n))$ and
$\pi_{\omega}(a^*(f_n))$ are $\sigma_t$-entire analytic
element\cite{BKP2}. Thus one can apply Theorem \ref{thm2.1} and
Corollary \ref{cor2.1} directly in this case.

(d) In applications to open systems\cite{Dav} and quantum
statistical mechanics\cite{BR}, one may need to construct a
Dirichlet form for a given $\{x_k : k\in I\}$, where each $x_k$ is
unbounded (self-adjoint) operator affiliated with $\cM$. By
employing appropriate approximation procedures, one may be able to
construct the Dirichlet form associated to $\{x_k : k\in
I\}$\cite{BK,BKP1} and then extend Theorem \ref{thm2.1} by
modifying the method used in this paper.
\end{remark}

\section{Proof of Theorem \ref{thm2.1}}

Before producing the proof of Theorem \ref{thm2.1}, we first
describe the basic ideas used in the proof, and then establish
necessary technical lemmas which will be needed in the proof.
Using the lemmas, we complete the proof at the last part of this
section.

 The inclusion $[\mathcal{Z}(\cM)\xi_0]\subset\mathcal{N}$ is easy to
 check. See the proof of Theorem \ref{thm2.1}. Thus we concentrate to
the inclusion $\mathcal{N}\subset[\mathcal{Z}(\cM)\xi_0]$. Note
that $\eta\in\mathcal{N}$ if and only if $\cE[\eta]=\langle
H^{1/2}\eta,H^{1/2}\eta\rangle=0.$ Since $\cE_k[\eta]\ge0$ for
$\eta\in\cH,\,\, k\in I,\,\, \eta\in\mathcal{N}$ if and only if
$\cE_k[\eta]=0$ for any $k\in I$. Since $f$ is an admissible
function or else $f=f_0$, it is easy to show that $\cE_k[\eta]=0$
if and only if
$$\|(\sigma_{t-i/4}(x_k)-j(\sigma_{t-i/4}(x_k)))\eta\|=0$$
for any $t\in\R$ and $k\in I$. See Lemma \ref{lemm3.1}. The above
implies that
\begin{equation}\label{3.1}
\|(\sigma_{-i/4}(x_k)-j(\sigma_{-i/4}(x_k)))\eta\|=0,\,\, k\in I.
\end{equation}
Suppose that $\eta$ is of the form
$\eta=\Delta^{1/4}Q\xi_0,\,Q\in\cM$. Then the above equality
implies that
$$[x_k,Q]\xi_0=0,\,\,\forall k\in I$$
and so
$$[x_k,Q]A'\xi_0=0,\,\,\forall k\in I$$
for any $A'\in\cM'$. Since $\cM'\xi_0$ is dense in $\cH$, we
conclude that $[x_k,Q]=0$ for $k\in I$. Since $\{x_k :k\in I\}$
generates $\cM$, $Q\in\cM'$. Thus $Q\in\cM\cap\cM'$ and
$\eta=\Delta^{1/4}Q\xi_0=Q\xi_0$. However, in general
$\eta\in\mathcal{N}$ can not be written as
$\eta=\Delta^{1/4}Q\xi_0,\,Q\in\cM$.

Note that $H$ is $J$-real, $JH=HJ$, and so
$J\mathcal{N}=\mathcal{N}.$ Any $\eta\in\cH$ can be written as
$\eta=\eta_r+i\eta_i$, where $\eta_r=(\eta+J\eta)/2$ and
$\eta_i=-i(\eta-J\eta)/2$. Thus $\eta\in\mathcal{N}$ implies
$\eta_r,\,\eta_i\in\cN$. Hence we may suppose that
$\eta\in\mathcal{N}\cap\cH^J$. Because of the Dirichlet property
\ref{2.1} (c) of $\cE[\eta]$, it can be shown that
$\eta\in\mathcal{N}\cap\cH^J$ implies
$\eta_+,\eta_-\in\mathcal{N}\cap\mathcal{P}$(Lemma \ref{lemm3.2}).
Thus the problem is reduced to the case
$\eta\in\mathcal{N}\cap\cP$.

Any $\eta\in D(\Delta^{-1/4})\cap\cP$ can be written as
$\eta=\Delta^{1/4}Q\xi_0$ where $Q$ is positive self-adjoint
operator affiliated with $\cM$(Lemma \ref{lemm3.4}). For any
$\eta\in
D(\Delta^{-1/4})\cap(\mathcal{N}\cap\cP),\,\eta=\Delta^{1/4}Q\xi_0$,
we use (\ref{3.1}) to show that
$$(x_kQ-Qx_k)\xi_0=0.$$
Using the facts that $\cM'\xi_0$ is dense in $\cH$ and that $\{x_k
:k\in I\}$ generates $\cM$, we will show that $Q$ is affiliated
with $\cM'$. Since $\Delta\xi=\xi$ for any
$\xi\in[\mathcal{Z}(\cM)\xi_0]$, we conclude that
$\eta\in[\mathcal{Z}(\cM)\xi_0]$. Next, we use the fact that
$D(\Delta^{-1/4})\cap(\mathcal{N}\cap\cP)$ is dense in
$(\mathcal{N}\cap\cP)$(Lemma \ref{lemm3.3}) to complete the proof
of Theorem \ref{thm2.1}.

Next, we collect technical lemmas which will be used in the
sequel. In the rest of this section, we assume that the conditions
in the Theorem hold.

 \begin{lemma}\label{lemm3.1} A vector $\eta\in\cH$ belongs to
 $\mathcal{N}$ if and only if the equality
 $$(\sigma_{t-i/4}(x_k)-j(\sigma_{t-i/4}(x_k)))\eta=0$$
 holds for any $t\in\mathbb{R}$ and $k\in I$.
\end{lemma}
\begin{proof}   Since
\begin{eqnarray*}
\langle\eta,H\eta\rangle&=&\cE(\eta,\eta)\\
                       &=&\sum_{k\in I}\cE_k(\eta,\eta),
\end{eqnarray*}
and $\cE_k(\eta,\eta)\ge0$ for $\eta\in\cH$ and $k\in I$,
$H\eta=0$ if and only if $\cE_k(\eta,\eta)=0$. Recall the
expression of $\cE_k(\eta,\eta)$ in (\ref{2.7}). Notice that
$f_0(t)>0$ for any $t\in\R$ by (\ref{2.4}). If $f$ is an
admissible function,  $f(t)\ge0$  by (\ref{2.3}) (a). Since $f$ is
analytic on a domain containing $I_{1/4}, f(t)>0$ except on a
countable set with no accumulation points. Thus the left hand side
of the expression in the lemma is zero except on a countable set
of $t\in\R$. Since $\sigma_{t-i/4}(x_k)$ is strongly continuous
with respect to $t\in\R$, we proved the lemma.
\end{proof}
 \begin{lemma}\label{lemm3.2} (a) $\mathcal{N}$ is a closed subspace of $\cH$.

 (b) $\Delta^{it}\mathcal{N}=\mathcal{N},\,\forall t\in\R.$

 (c) $J\mathcal{N}=\mathcal{N}.$

 (d) $\eta\in\mathcal{N}\cap\cH^J$ implies $\eta_+,\,\eta_-\in\mathcal{N}\cap\cP.$
\end{lemma}
\begin{proof}   (a) Since $H$ is self-adjoint (closed), (a) is
obvious.

(b) Notice that
$$H\Delta^{-is}\eta=\Delta^{-is}(\Delta^{is}H\Delta^{-is})\eta$$
and $\Delta^{is}H\Delta^{-is}$ is the Dirichlet operator
associated to the Dirichlet form constructed with $\{\sigma_s(x_k)
: k\in I\}$. Note that
$\sigma_{t-i/4}(\sigma_s(x_k))=\sigma_{t+s-i/4}(x_k)$. Thus, if
$\eta \in\mathcal{N},\,\Delta^{-is}\eta\in\cN$ by Lemma
\ref{lemm3.1}. Hence $\Delta^{-is}\cN\subset\cN$ for any $s\in\R$,
which also implies $\cN\subset\Delta^{is}\cN$ for any $s\in\R$.

(c) Since each $\cE_k$ is $J$-real (Theorem 2.1 (b) of
\cite{Par2}), it is easy to check that $H$ is $J$-real, $JH=HJ.$
Thus $HJ\eta=JH\eta=0$ if $\eta\in\cN$ and so $J\cN\subset\cN$.
Since $J^2=\1$, we also have that $\cN\subset J\cN$.

(d) Let $\eta\in\cN\cap\cH^J$, and $\eta=\eta_+-\eta_-.$ Notice
that
\begin{eqnarray*}\nonumber
0&=&\cE(\eta,\eta)\\
  &=&\cE(\eta_+,\eta_+)-2\cE(\eta_+,\eta_-)+\cE(\eta_-,\eta_-).
\end{eqnarray*}
Here we have used the fact that $\eta\in D(\cE)$ implies
$\eta_+,\eta_-\in D(\cE)$. Since $\cE(\eta_+,\eta_-)\le0$ by
(\ref{2.1}) (c) (Theorem  2.1 (c) of \cite{Par2}), we have that
$\cE(\eta_+,\eta_+)=\cE(\eta_-,\eta_-)=0$, which imply
$H\eta_+=H\eta_-=0.$
\end{proof}

\begin{lemma}\label{lemm3.3} (a) For any bounded, positive definite, continuous function $f : \R \rightarrow \R$,
                      $$f(\log \Delta)(\cN\cap\cP)\subset\cN\cap\cP.$$
 (b) $(\bigcap_{\alpha\in\R}D(\Delta^{\alpha}))\cap(\cN\cap\cP)$ is dense in $\cN\cap\cP$.
 \end{lemma}
\begin{proof}   (a) Let $f$ be a bounded, positive definite, continuous function on $\R$. Then $f$ can be written as
                        $$f(x)=\int e^{itx}d\mu(t),$$
                        where $\mu$ is a positive finite Borel measure on $\R$. Thus
                           $$f(\log \Delta)=\int \Delta^{it}d\mu(t).$$
                        The inclusion
                      $$f(\log \Delta)\cP\subset\cP$$
         holds by the fact that $\Delta^{it}\cP\subset\cP$(Proposition 2.5.26 of \cite{BR}).
         Due to Lemma \ref{lemm3.2} (b), the inclusion
         $$f(\log \Delta)\cN\subset\cN$$
         also holds. This proved the part (a) of the lemma.

(b) Let $$f_n(x):=e^{-x^2/2n^2}.$$ Then by the part (a) of the
 lemma, $$f_n(\log \Delta)(\cN\cap\cP)\subset\cN\cap\cP.$$
 For any $\eta\in \cN\cap\cP$, $f_n(\log\Delta)\eta\in
 D(\Delta^\alpha)$ for any $\alpha\in\R$, and
 $f_n(\log\Delta)\eta\rightarrow\eta$ as $n\rightarrow\infty$.
 This proved the part (b).
\end{proof}

\begin{lemma}\label{lemm3.4}
Let $\eta\in D(\Delta^{-1/4})\cap\cP$. Then there is a positive
self-adjoint operator $Q$ affiliated with $\cM$ such that
$Q\xi_0\in D(\Delta^{1/4})$ and $\eta=\Delta^{1/4}Q\xi_0$.
\end{lemma}
\begin{proof}   We use the method similar to that employed in the
proof of Proposition 2.5.27(1)of \cite{BR}. Let $\eta\in
D(\Delta^{-1/4})\cap\cP$. For any $A\in\cM$,
$\Delta^{-1/4}j(A^*)j(A)\xi_0\in\cP$ and so
$$\langle\Delta^{-1/4}j(A^*)j(A)\xi_0,\eta\rangle\ge0,\,\,\forall A\in\cM,$$
which implies
$$\langle j(A^*)j(A)\xi_0,\Delta^{-1/4}\eta\rangle\ge0,\,\,\forall A\in\cM.$$
Define an operator $\tilde{Q},\,D(\tilde{Q})=\cM'\xi_0$, by
$$\tilde{Q}j(B)\xi_0=j(B)\Delta^{-1/4}\eta,\,\,\forall B\in\cM.$$
Then for any unitary $U'\in\cM'$,
$$U'\tilde{Q}j(B)\xi_0=U'j(B)\Delta^{-1/4}\eta=\tilde{Q}U'j(B)\xi_0,$$
and so
$$U'^*\tilde{Q}U'=\tilde{Q}.$$
For any $A\in\cM$,
\begin{eqnarray*}
\langle j(A)\xi_0,\tilde{Q}j(A)\xi_0\rangle&=&\langle
j(A)\xi_0,j(A)\Delta^{-1/4}\eta\rangle\\
&=&\langle\Delta^{-1/4}j(A^*)j(A)\xi_0,\eta\rangle\\
&\ge&0.
\end{eqnarray*}
Thus $\tilde{Q}$ is a positive symmetric operator. Notice that for
any unitary $U'\in\cM'$, $U'D(\tilde{Q})\subset D(\tilde{Q})$. Let
$Q$ be the Friedrichs extension of $\tilde{Q}$. By the uniqueness
of Friedrichs extension
$$U'^*QU'=Q$$
for any unitary $U'\in\cM'$. Thus $Q$ is affiliated with $\cM$.
Since $Q\xi_0=\Delta^{-1/4}\eta,\,Q\xi_0\in D(\Delta^{1/4})$ and
$\eta=\Delta^{1/4}Q\xi_0$.
\end{proof}

\begin{lemma}\label{lemm3.5}
Let $\eta\in D(\Delta^{-1/4})\cap(\cN\cap\cP)$ and
$\eta=\Delta^{1/4}Q\xi_0$ as in Lemma \ref{lemm3.4}. Then
$x_k\xi_0\in D(Q)$ and $x_kQ\xi_0=Qx_k\xi_0$ for any $k\in I$.
\end{lemma}
\begin{proof}   Since $\eta\in\cN$, it follows from Lemma \ref{lemm3.1} that
$$[\sigma_{-i/4}(x_k)-j(\sigma_{-i/4}(x_k))]\Delta^{1/4}Q\xi_0=0,\,\,\,k\in I.$$
Recall that $\cM_0$ is the dense subset of $\cM$ consisting of
$\sigma_t$-entire analytic elements. For any $A\in\cM_0$,
\begin{eqnarray}\label{3.2}
0&=&\langle
\sigma_{i/4}(A)\xi_0,[\sigma_{-i/4}(x_k)-j(\sigma_{-i/4}(x_k))])\Delta^{1/4}Q\xi_0\rangle\nonumber\\
&=&\langle \sigma_{i/4}(x_k)\sigma_{i/4}(A)\xi_0,\Delta^{1/4}Q\xi_0\rangle\\
 &&-\langle j(\sigma_{i/4}(x_k))\sigma_{i/4}(A)\xi_0,\Delta^{1/4}Q\xi_0\rangle.\nonumber
\end{eqnarray}
For any $A\in\cM_0$,
\begin{eqnarray}\label{3.3}
\sigma_{i/4}(x_k)\sigma_{i/4}(A)\xi_0&=&\sigma_{i/4}(x_kA)\xi_0\nonumber\\
                                     &=&\Delta^{-1/4}x_kA\xi_0
\end{eqnarray}
and
\begin{eqnarray}\label{3.4}
j(\sigma_{i/4}(x_k))\sigma_{i/4}(A)\xi_0&=&j(\sigma_{i/4}(x_k))j(\sigma_{-3i/4}(A^*))\xi_0\nonumber\\
                                     &=&\Delta^{-1/4}j(\sigma_{i/2}(x_k))j(\sigma_{-i/2}(A))\xi_0 \nonumber\\
                                     &=&\Delta^{-1/4}j(\sigma_{i/2}(x_k))A\xi_0.
                                     \end{eqnarray}
 Substituting (\ref{3.3}) and (\ref{3.4}) into (\ref{3.2}), we
 have $$\langle A\xi_0,[x_k-j(\sigma_{-i/2}(x_k))]Q\xi_0\rangle=0$$
 for any $A\in\cM_0$ and $k\in I$. Since $\cM_0\xi_0$ is dense in
 $\cH$,
 $$[x_k-j(\sigma_{-i/2}(x_k))]Q\xi_0=0.$$
 Since $Q$ is affiliated with $\cM$,
 $j(\sigma_{-i/2}(x_k))Q\xi_0=Qj(\sigma_{-i/2}(x_k))\xi_0=Qx_k\xi_0$,
 and so $x_k\xi_0\in D(Q)$ and $x_kQ\xi_0=Qx_k\xi_0$.
\end{proof}
Next, we use Lemma \ref{lemm3.4} and Lemma \ref{lemm3.5} to the
prove the following result:
\begin{proposition}\label{prop3.1}
Let $\eta\in D(\Delta^{-1/4})\cap(\cN\cap\cP)$. Then there is a
positive self-adjoint operator $Q$ affiliated with
$\mathcal{Z}(\cM)$ such that $\eta=Q\xi_0$.
\end{proposition}
\begin{proof} Let $\eta=\Delta^{1/4}Q\xi_0$ as in Lemma
\ref{lemm3.4}. Due to Lemma \ref{lemm3.5},
$$x_kQ\xi_0=Qx_k\xi_0,\,\,\forall k\in I.$$ Since
$x_k\in\cM_{1/2}\subset\cM$ and $Q$ is affiliated with $\cM$,
\begin{eqnarray*}
x_kQj(A)\xi_0&=&j(A)x_kQ\xi_0\\
             &=&j(A)Qx_k\xi_0\\
             &=&Qx_kj(A)\xi_0
\end{eqnarray*}
for any $A\in\cM$, and so
$$x_kQj(A)\xi_0=Qx_kj(A)\xi_0,\,\,\forall A\in\cM.$$ Notice that
for any $x_{k_1},\,x_{k_2}\in\{x_k : k\in I\}$
\begin{eqnarray}\label{3.5}
x_{k_1}x_{k_2}Qj(A)\xi_0&=&x_{k_1}Qx_{k_2}j(A)\xi_0\nonumber\\
                       &=&x_{k_1}Qj(A)j(\sigma_{-i/2}(x_{k_2}))\xi_0\nonumber\\
                       &=&Qx_{k_1}j(A)j(\sigma_{-i/2}(x_{k_2}))\xi_0\nonumber\\
                       &=&Qx_{k_1}x_{k_2}j(A)\xi_0.
                       \end{eqnarray}
                       Let $\tilde{\cM}$ be the $*$-algebra
                       generated by $\{x_k : k\in I\}$. Then
                       $\tilde{\cM}$ is dense in $\cM$ by the
                       assumption in Theorem \ref{thm2.1}. The
                       relation (\ref{3.5}) implies that for any
                       $x\in\tilde{\cM}$,
                       $$xQj(A)\xi_0=Qxj(A)\xi_0,\,\,\forall A\in\cM.$$
  For given $x\in\cM$, choose a sequence $x_n\in\tilde{\cM}$ such that
  $x_n\rightarrow x$ strongly. Then
  \begin{eqnarray*}
Qx_nj(A)\xi_0&=&x_nQj(A)\xi_0\\
             &\rightarrow& xQj(A)\xi_0\quad\text{as}\quad n\rightarrow \infty.
             \end{eqnarray*}
 Due to the closedness of $Q$ and the fact that
 $x_nj(A)\xi_0\rightarrow xj(A)\xi_0$ as $n\rightarrow\infty$, we
 conclude that $xj(A)\xi_0\in D(Q)$ and
 \begin{equation}\label{3.6}
 xQj(A)\xi_0=Qxj(A)\xi_0
 \end{equation}
 for any $x,A\in\cM$.

 Denote by
 $$(\cM\times\cM')\xi_0:=\{AA'\xi_0 : A\in\cM,\,A'\in\cM'\}.$$
By (\ref{3.6}), $(\cM\times\cM')\xi_0\in D(Q)$ and for any
$A_1,\,A_2\in\cM$ and $A'_1,\,A'_2\in\cM'$
\begin{eqnarray}\label{3.7}
A_1QA_2A_2'\xi_0&=&A_1A_2QA'_2\xi_0\nonumber\\
             &=&QA_1A_2A'_2\xi_0
 \end{eqnarray}
 and
 \begin{eqnarray}\label{3.8}
A'_1QA_2A_2'\xi_0&=&QA_2A'_1A'_2\xi_0\nonumber\\
             &=&QA'_1A_2A'_2\xi_0.
 \end{eqnarray}
 Let $Q_0$ be the restriction of $Q$ on $(\cM\times\cM')\xi_0$.
 Then $Q_0$ is a positive symmetric operator. It follows from
 (\ref{3.7}) and (\ref{3.8}) that for any unitary
 $U\in\cM,\,U'\in\cM'$,
  \begin{eqnarray}\label{3.9}
&&U^*Q_0U=Q_0,\\ &&{U'}^*Q_0U'=Q_0.\nonumber
 \end{eqnarray}
 Notice that $U$ and $U'$ leave $(\cM\times\cM')\xi_0$ invariant.
  Let $\hat{Q}$ be the Friedrichs of $Q_0$. By the uniqueness of
  Friedrichs extension,
  \begin{eqnarray*}
&&U^*\hat{Q}U=\hat{Q},\\ &&{U'}^*\hat{Q}U'=\hat{Q}.\nonumber
 \end{eqnarray*}
 for any unitary $U\in\cM,\,\,U'\in\cM'$. Thus $\hat{Q}$ is
 affiliated with $\mathcal{Z}(\cM)$. By the inclusions
 $\cM'\xi_0\subset(\cM\times\cM')\xi_0\subset D(Q)$ and the uniqueness of
 the Friedrichs extension, $\hat{Q}=Q.$ Since $\Delta\xi=\xi$ for
any $\xi\in[\mathcal{Z}(\cM)\xi_0]$,
$\eta=\Delta^{1/4}Q\xi=Q\xi_0.$ This completes the proof of the
proposition.
\end{proof}

We are ready to prove Theorem \ref{thm2.1}.

\vspace*{0.3cm} \noindent {\bf Proof of Theorem \ref{thm2.1}.}
\quad The inclusion
\begin{equation}\label{3.10}
[\mathcal{Z}(\cM)\xi_0]\subset\cN
\end{equation}
is easy to prove as follow: Let$\xi\in\mathcal{Z}(\cM)\xi_0$. Then
$\xi=A\xi_0$ for some $A\in\mathcal{Z}(\cM).$ Thus
 \begin{eqnarray*}
&&[\sigma_{t-i/4}(x_k)-j(\sigma_{t-i/4}(x_k))]A\xi_0\\
&&=A[\sigma_{t-i/4}(x_k)-j(\sigma_{t-i/4}(x_k))]\xi_0\\ &&=0.
 \end{eqnarray*}
 By Lemma \ref{lemm3.1}, $\xi\in\cN$. Since $\cN$ is closed by
 Lemma \ref{lemm3.2} (a), the closure of $\mathcal{Z}(\cM)\xi_0$
 is a subspace of $\cN$. This proved the inclusion (\ref{3.10}).

 Next, we prove the inclusion
 \begin{equation}\label{3.11}
\cN\subset[\mathcal{Z}(\cM)\xi_0].
\end{equation}
Any $\eta\in\cN$ can be written as $\eta=\eta_r+i\eta_i$, where
$\eta_r=(\eta+J\eta)/2$ and $\eta_i=-i(\eta-J\eta)/2.$ By Lemma
\ref{lemm3.2} (c), $\eta_r,\,\eta_i\in\cN.$ Note that
$\|\eta\|^2=\|\eta_r\|^2+\|\eta_i\|^2.$ Thus we may assume that
$\eta$ is $J$-real, $\eta\in\cN\cap\cH^J$. $\eta$ is decomposed
uniquely as $\eta=\eta_+-\eta_-,\,\eta_+,\,\eta_-\in\cP$ and
$\eta_+\perp \eta_-.$ See Proposition 2.5.28 (3) of \cite{BR}. By
Lemma \ref{lemm3.2} (d), $\eta_+,\,\eta_-\in\cN\cap\cP$. Lemma
\ref{lemm3.3} (b) implies that $D(\Delta^{-1/4})\cap(\cN\cap\cP)$
is dense in $\cN\cap\cP$. Thus Lemma \ref{lemm3.3} (b) and
Proposition \ref{prop3.1} imply that
$\eta_+,\,\eta_-\in[\mathcal{Z}(\cM)\xi_0]$, and so
$\eta\in[\mathcal{Z}(\cM)\xi_0]$. This completes the proof of
Theorem \ref{thm2.1}.\quad$\square$

\section{Ergodicity of Markovian Semigroups for Quantum Spin
Systems}

 In this section, we first describe the translation  invariant
 Markovian semigroups for quantum spin systems constructed in
 \cite{Par1}, and then apply Theorem \ref{thm2.1} (and Corollary
 \ref{cor2.1}) to show the ergodicity of the semigroups in region
 of high temperatures where the uniqueness of KMS-state holds.

 Let us describe quantum spin systems briefly. For details, we
 refer to Section 6.2 of \cite{BR}. Let $\mathbb{Z}^d$ be a
 $d-$dimensional lattice space and let $\cF$ denote the family
 of all finite subsets of $\mathbb{Z}^d$.
 Let $\cA$ be a $C^*$-algebra with norm $\|\cdot\|$ defined as the
 inductive limit over a finite-dimensional matrix algebra
 $\mathbb{M}$. For any $X\in\cF$, let $\cA_X$ denote the subalgebra
 localized in $X$, i.e., the subalgebra in $\cA$ isomorphic to $\mathbb{M}^X$.
  An element $A\in\cA$ will be called {\it local} if there is some
  $Y\in\cF$ such that $A\in\cA_Y$. By $\cA_0$ we denote the
  subset of all local elements, i.e.,
  $\cA_0=\bigcup_{X\in\cF}\cA_X.$

  Let $\Phi:=\{\Phi_X\}_{X\in\cF}$ be an interaction, i.e., a
  family of self-adjoint element in $\cA$. Suppose that
 \begin{equation}\label{4.1}
 \|\Phi\|_\lambda :=\sup_{i\in\mathbb{Z}^d}\sum_{X\in\cF\, :\, i\in X}
 e^{\lambda|X|}\|\Phi_X\|<\infty
  \end{equation}
  for some $\lambda>0$, where $|X|=\text{card}(X).$ Define a
  derivation $\delta$ by
\begin{eqnarray}\label{4.2}
&&D(\delta)=\cA_0\nonumber, \\ &&
\delta(A)=-i\sum_{X\cap\Lambda\neq\emptyset}[\Phi,A],\,\,A\in\cA_{\Lambda}.
\end{eqnarray}
Then $\cA_0$ is a norm-dense $*$-subalgebra of analytic element of
the closure $\overline{\delta}$ of $\delta$. Thus
$\overline{\delta}$ generates one-parameter group of
$*$-automorphism $\tau$ of $\cA$. Let $\omega$ be a $\tau$-KMS
state corresponding to the interaction $\Phi$.

 Let $(\cH_{\omega}, \pi_{\omega}, \Omega_{\omega})$ be the GNS
 representation of $(\cA,\omega)$. For the standard form, we
 choose $\cH=\cH_\omega , \cM=\pi_\omega(\cA)''$ and
 $\xi_0=\Omega_\omega.$ By the uniqueness of the modular
 automorphism(see Theorem 5.3.10 of \cite{BR}), one may identify
 $\sigma_t=\tau_t,\,t\in\R$, on $\cM$. In this section, we denote
 by $\cM_0$ the algebra of local elements, i.e.,
 $\cM_0=\pi_{\omega}(\cA_0)$. Every element $A\in\cM_0$ is an
 analytic element for $\sigma_t$. For a given $\lambda>0$, put
 $\gamma=\lambda/2\|\Phi\|_\lambda.$ Then for any
 $s\in(-\gamma,\gamma)$ the series
 \begin{equation}\label{4.3}
 \sigma_{is}(A)=\sum_{n=0}^\infty\frac{(-is)^n}{n!}\delta^n(A),\,\,A\in\cM_0,
 \end{equation}
 converges absolutely, where $\delta$ is the derivation given
 (\ref{4.2}). See the proof of Theorem 5.2.4 of \cite{BR}. From
 now on, we assume that $\Phi$ is chosen sufficiently small so
 that $\gamma>1/2.$

 We now turn to Dirichlet form for quantum spin sytems\cite{Par1}.
 Let $\{\tau_j\}_{j\in\mathbb{Z}^d}$ be the translational
 automorphism on $\cM$ corresponding to the translation of the lattice
 by vectors $j\in\mathbb{Z}^d$. Let
 $x^a\in\pi_\omega(\mathbb{M}),\,a=1,2,...,D$, be a basis
 of $\pi_\omega(\mathbb{M})$ consisting of self-adjoint elements of norm one and let
 $x_j^a=\tau_j(x^a),\,j\in\mathbb{Z}^d.$ For the family
$\{x_j^a :j\in\mathbb{Z}^d, a=1,2,...,D\}$ and an admissible
function $f$ (or else $f_0$), let $(\cE,D(\cE))$ be the quadratic
form defined as in (\ref{2.5}) - (\ref{2.7}):
\begin{eqnarray}\label{4.5}
&&D(\cE)=\{\xi\in\cH : \sum_{j\in\mathbb{Z}^d}
\sum_{a=1}^D\cE_{a,j}[\xi]<\infty\},\nonumber\\ &&
\cE[\xi]=\sum_{j\in\mathbb{Z}^d}
\sum_{a=1}^D\cE_{a,j}[\xi],\,\,\xi\in D(\cE)
\end{eqnarray}
where
\begin{equation}\label{4.6}
\cE_{a,j}[\xi]=\int\|(\sigma_{t-i/4}(x^a_j)-j(\sigma_{t-i/4}(x^a_j)))\xi\|^2f(t)dt.
\end{equation}
The following is Theorem 5.1 of \cite{Par1}:

\begin{theorem}\label{thm4.1}: {\bf (Theorem 5.1 of \cite{Par1})} Let
$f$ be an admissible function such that $p$ in (\ref{2.3}) (c) is
greater than $d+1$, i.e., $p>d+1$. Let the interaction $\Phi$ be
of finite range and translation invariant. Then the form
$(\cE,D(\cE))$ defined as in (\ref{4.5}) - (\ref{4.6}) is a
densely defined Dirichlet form which generates a translation
invariant, symmetric, Markovian semigroup.
\end{theorem}
\begin{remark}\label{rem4.1}
The strongly decay property of $f$, i.e., $p>d+1$, has been used
to show that $D(\cE)$ is dense in $\cH$. See the proof of Theorem
5.1 of \cite{Par1}. The function $f_0$ given in (\ref{2.4}) decays
exponentially fast and so the conclusion in Theorem \ref{thm4.1}
holds for $f=f_0$.
\end{remark}
In order to describe the main result, we need to replace $\Phi$ by
$\beta\Phi$, where $\beta$ is the inverse temperature. Then the
condition $\gamma>1/2$ is equivalent to
$(\lambda/2\beta\|\Phi\|_\lambda)>\frac12$. This is,
$\beta\|\Phi\|_\lambda<\lambda.$ The following is the main result
in this section:

\begin{theorem}\label{thm4.2} Let $f$ be either
an admissible function satisfying the decay property in Theorem
\ref{thm4.1} or else $f=f_0$. Let the interaction $\Phi$ be of
finite range and translation invariant. For $\{x^a_j :
j\in{\mathbb{Z}^d}, a=1,2,...,D\}$, let $\{T_t\}_{t\ge0}$ be the
translation invariant Markovian semigroup associated to the
Dirichlet form defined as in (\ref{4.5})-(\ref{4.6}). Assume that
$\beta\|\Phi\|_\lambda$ is sufficiently small so that
$(\tau,\beta)$-KMS state for $\Phi$ is unique. Then the Markovian
semigroup $\{T_t\}_{t\ge0}$ is ergodic.
\end{theorem}
\begin{remark}\label{rem4.2}
The region of high temperatures where the uniqueness of
$(\tau,\beta)$-KMS state holds can be given explicitly. For an
instance, see Proposition 6.2.25 of \cite{BR}. For one-dimensional
models with uniform bounded surface energies,  the uniqueness of
$(\tau,\beta)$-KMS state is independent of temperature(Theorem
6.2.47 of \cite{BR}). However, we still need the condition
$\beta\|\Phi\|_\lambda<\lambda.$
\end{remark}

\vspace*{0.3cm} \noindent {\bf Proof of Theorem \ref{thm4.2}.} By
the condition $\beta\|\Phi\|_\lambda<\lambda$, the series
(\ref{4.3}) converges absolutely on a region containing
$[-\beta/2,\beta/2]$. Thus it is easy to see that $\{x^a_j :
j\in\mathbb{Z}^d, \, a=1,2,...,D\}\subset\cM_{\beta/2}$. Since the
$*$-algebra generated by the family is $\cM_0$, which is dense in
$\cM$, the condition in Theorem \ref{thm2.1} hold. The uniqueness
of the $(\tau,\beta)$-KMS state $\omega$ implies that $\omega$ ia
an extremal $(\tau,\beta)$-KMS state, and hence a factor state by
Theorem 5.3.30 of \cite{BR}. Thus $\cM$ is a factor, and so
$\{T_t\}_{t\ge0}$ is ergodic by Corollary \ref{cor2.1}.
 $\quad \square$.

%\begin{acknowledgements}
\vspace{0.2cm} \noindent {\bf Acknowledgements}:  This work was
supported by Korea Research Foundation (KRF-2003-005-C00010),
Korean Ministry of Education.
%\end{acknowledgements}


\begin{thebibliography}{99}
\bibitem[Acc]{Acc} L. Accardi, Topics in quantum probability, {\it Phys.Rep.}
{\bf 77}, 169-192 (1981).
\bibitem[Ara]{Ara} H. Araki, Some properties of modular conjugation operator of von Neumann
algebras ans noncommutative Radon-Nikodym theorem with chain rule,
{\it Pacific J. Math.} {\bf 50} (2), 309-354 (1974).
\bibitem[BK]{BK}C. Bahn and C. K. Ko, Construction of unbounded Dirichlet forms on standard forms of von Neumann
Algebras, {\it J. Korean Math. Soc.} {\bf 39} (6) 931-951 (2002).
\bibitem[BKP1]{BKP1} C. Bahn, C. K. Ko amd Y. M. Park, Dirichlet forms
and symmetric Markovian semigroups on CCR Algebras with quasi-free
states, {\it J. Math. Phys.}, {\bf 44}, 723-753 (2003).
\bibitem[BKP2]{BKP2} C. Bahn, C. K. Ko amd Y. M. Park, Dirichlet forms
and symmetric Markovian semigroups on $\Z_2$-graded von Neumann
algebras, {\it Rev. Math. Phys.}, {\bf 15}, 823-845 (2003).
\bibitem[BR]{BR} O.  Bratteli and  D. W.  Robinson,
 {\it  Operator algebras  and quantum  statistical mechanics},
 Springer-Verlag, New  York-Hei\-delberg-Berlin, vol I 1979, vol. II (1981).
\bibitem[Cip1]{Cip1} F. Cipriani, Dirichlet forms  and Markovian semigroups on standard
forms of von Neumann algebras,  {\it J. Funct. Anal.} {\bf 147},
259-300 (1997).
\bibitem[Cip2]{Cip2} F. Cipriani, Perron Theory for Positive Maps Semigroups on von Neumann
Algebras, {\it Canadian Math. Soc., Conference Proceedings}, Vol.
2a, 115-123 (2000).
\bibitem[CFL]{CFL} F. Cipriani, F. Fagnola and J. M. Lindsay, Spectral Analysis and
Feller Properties for Quantum Ornstein-Uhlenbeck Semigroups, {\it
Comm. Math. Phys.} {\bf 210},  85-105 (2000).
\bibitem[Dav]{Dav} E. B. Davies, {\it Quantum theory of open systems},
 Academic Press, London-New York-San Francisco, (1976).
\bibitem[GL1]{GL1} S. Goldstein and J. M. Lindsay,
KMS-symmetric Markov semigroups, {\it Math. Zeit.} {\bf 219},
590-608 (1995).
\bibitem[GL2]{GL2} S. Goldstein and J. M. Lindsay, Markov semigroups KMS-symmetric
for a weight, {\it Math. Ann.} {\bf 313}, 39-67 (1999).
\bibitem[KP]{KP} C. K. Ko and Y. M. Park, Construction of a Family of Quantum
Ornstein-Uhlenbeck Semigroups, {\it J. Math. Phys.}, {\bf 45},
609-627 (2004).
\bibitem[MZ1]{MZ1} A. W.  Majewski and B. Zegarlinski, Quantum stochastic  dynamics I:
Spin systems on a  lattice, {\it MPEJ} {\bf 1}, Paper  2 (1995).
\bibitem[MZ2]{MZ2} A. W. Majewski and B. Zegarlinski, Quantum stochastic dynamics II,
 {\it Rev. Math. Phys.} {\bf 8} (5), 689-713 (1996).
\bibitem[Par1]{Par1} Y. M. Park, Construction of Dirichlet forms on standard
forms of von Neumann algebras,  {\it Infinite Dimensional
Analysis, Quantum Probability and Related Topics}, Vol. 3, No. 1,
1-14 (2000).
\bibitem[Par2]{Par2} Y. M. Park, Remark on the Structure of Dirichlet Forms
on the Standard Forms of von Neumann Algebras, arXiv.
Math-ph/04001, to be appeared in {\it Infinite Dimensional
Analysis, Quantum Probability and Related Topics}.
\bibitem[Part]{Part} K. R. Parthasarathy, {\it An introduction to quantum
stochastic calculus}, Birkh\"auser, Basel (1992).
\bibitem[RS]{RS} M.Reed and B.Simon, {\it Method of modern mathmatical
physics I, II}, Academic press (1980).
\bibitem[Zeg]{Zeg} B. Zegarlinski, Analysis of Classical and Quantum Interacting Partical Systems,
In {\it Quantum Probability and White Noise Analysis}, Vol. XIV,
eds, L. Accard and F. Fagnola, World Scientific, 241-336(2000).
\end{thebibliography}
\end{document}